\documentclass[%
superscriptaddress,
preprint,
showpacs,preprintnumbers,
amsmath,amssymb,
aps,
prb,
]{revtex4-1}
\usepackage{graphicx}
\usepackage{dcolumn}
\usepackage{bm}
\catcode`\"=13
\def"#1{\v #1}

\begin{document}


\title{Physical origin underlying the entropy loss upon hydrophobic hydration}

\author{Alja"z Godec}
\email{aljaz.godec@ki.si}
\author{Franci Merzel}
\email{franci.merzel@ki.si}
\affiliation{National Institute of Chemistry, Hajdrihova 19,1000 Ljubljana, Slovenia}



\date{\today}
\maketitle

\noindent 
{\bf The hydrophobic
effect (HE) is commonly associated with the demixing of oil and water at
ambient conditions and plays the leading role in determining the
structure and stability of biomolecular assembly in aqueous solutions.
On the molecular scale HE has an entropic origin. It is
believed that hydrophobic particles induce order in the
surrounding water by reducing the volume of
configuration space available for hydrogen bonding. 

Here we show with computer simulation results that this
traditional picture is not correct. Analyzing collective fluctuations in
water clusters we are able to provide a fundamentally new
picture of HE based on  
pronounced many-body correlations affecting the switching of hydrogen
bonds between molecules. These correlations emerge as a non-local 
compensation of reduced fluctuations of local electrostatic
fields in the presence of an apolar solute.}

The HE has a multifaceted nature, \textit{i.e.} its physical
manifestation depends on the length-scale \cite{Chand}.
On the mesoscale, \textit{i.e.} hydration of an
assembly of hydrophobic units or an extended hydrophobic surface, HE
is driven by energy/enthalpy and occurs as a 'dewetting' transition
\cite{ChanJCPB,Raj,Mitt} which 
has far-reaching consequences for processes such as
protein folding \cite{HuangPNAS,Papoian} and nanoparticle
self-assembly \cite{Rabani}. 
Meanwhile, HE on the molecular scale has
an entropic origin \cite{Still,Chand}, particularly near room temperature
and lower, while it is believed to eventually become energy/enthalpy
driven at higher temperatures \cite{Murthy,LynBell}. Furthermore, the molecular
scale hydration thermodynamics (hydrophobe solubilities, partitioning
of hydration free energy into
energy/enthalpy and entropy contributions, etc.) appear to be well
established and can be worked out, for example, using scaled-particle
theory \cite{PrattRMP} or information theory \cite{Humm}. While those
theories are successful in predicting solubilities of hydrophobic
solutes and several related thermodynamic features, they do not
provide deeper 
insight into the physical mechanism underlying HE. 
Notwithstanding all efforts and advances in the field
\cite{Tanf,Still,BenN,PrattAR,Chand,PrattRMP,ChanJCPB,Raj,Mitt} the 
physical picture of HE is still far from being complete and even
fundamental issues, such as the mechanism underlying hydrophobicity
on different length scales, still have to be clarified.
From the physical
point of view the most puzzling feature of HE remains the microscopic
picture of entropy loss upon hydrophobic hydration. 

Intuitively, the
entropy loss is usually attributed to the reduction of
configuration space available for hydrogen-bonding \cite{Still,Chand},
which is due to the fact that water molecules need to reorganize
around a hydrophobic solute to avoid sacrificing hydrogen bonds. This
is supposed to lead to remnants of clathrate structures \cite{Bow}, which are,
however, not rigid and their quantitative importance for understanding
hydrophobicity remains questionable \cite{PrattAR}. Furthermore, even the
actual role of hydrogen bonds for the HE is apparently not entirely clear
\cite{LynBell,Chaterjee}.

Thus, the physically
most intriguing question to be answered still remains:
If HE is entropy driven, what specifically causes the loss of
entropy? Since entropy loss is directly related to a reduction of
available volume in configuration space, how does it affect
degrees of freedom of water molecules?

Here we present conclusive simulation results which unravel a fundamentally new
picture of the mechanism of HE based on 
pronounced many-body correlations affecting the intermolecular
exchanging of hydrogen bonds. We carry out constant pressure Monte-Carlo
simulations of TIP5P water \cite{Jorg} and model hard sphere solutes in an
orthogonal simulation box with periodic boundary conditions (see Methods section for
details). 

\begin{figure}
\includegraphics[width=0.5\textwidth]{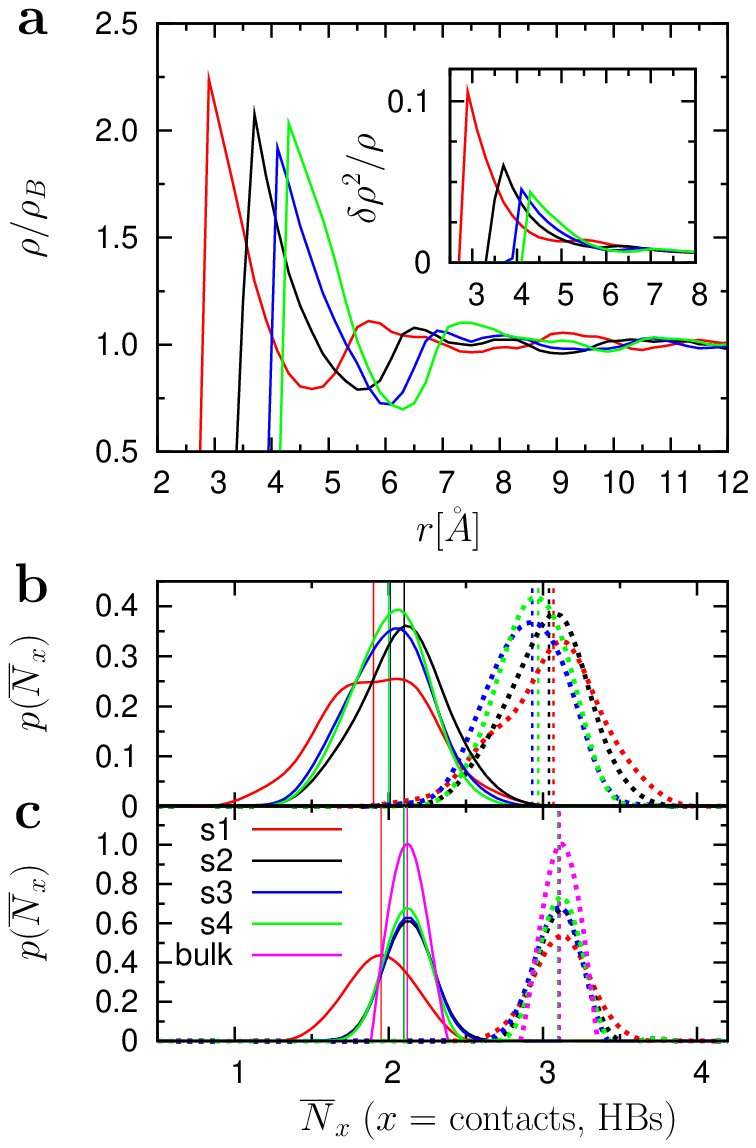}
\caption{}
\label{fg1}
\end{figure}

\begin{figure}
\includegraphics[width=0.85\textwidth]{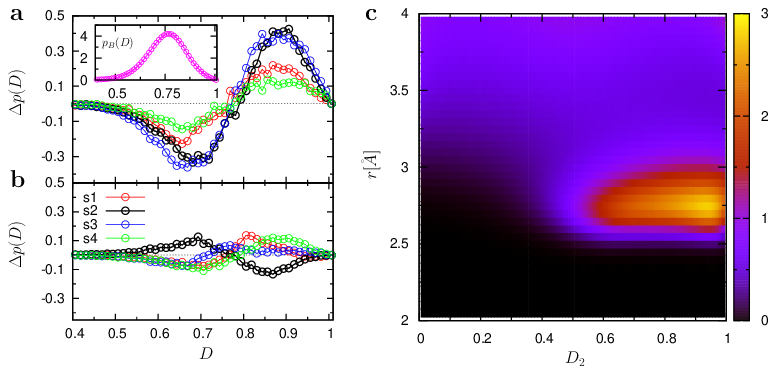}
\caption{}
\label{fg2}
\end{figure}

In order to present the conceptual change in our understanding of the
HE we first address the inconsistencies of the traditional picture.
The radial correlation function, $\frac{\rho(r)}{\rho_B}\equiv
g(r)=\rho_B^{-1}\langle \sum_{i=1}^N \delta 
(r-|\bm{r}_i|) \rangle$, is used to quantify the degree of
translational ordering of water molecules around hydrophobic solutes. 
We find a non-monotonic dependence of the contact density on particle size,
which is due to commensurability of the solute surface and water
packing. The relative density
fluctuations monotonically decrease with solute size. 
Two water molecules are defined to be in close contact if their
intermolecular distance is less than 3 \AA\ , and they are said to be 
hydrogen-bonded if they are in close contact and if the 
angle $O-H\cdots O$ is larger than $150^{\circ}$. The cutoff
distance for neighbors in close contact is set at 3 \AA\ and is more
appropriate with 
respect to the conventional definition of 
3.5 \AA\ \cite{Laage1}, as there is no preferential
mutual orientation beyond the distance of 3 \AA\ (see Fig. \ref{fg2}c).
The distribution of the number of water
molecules in close contact and the number of hydrogen bonded contacts
per water molecule located in the first and second 
hydration shell and in bulk water is shown in Fig. \ref{fg1}b and
c. Except for the smallest solute (with radius 1.4 \AA) there is no
appreciable difference (say of the order of $\ge$ 0.5) in the number
of total and hydrogen-bonded 
contacts with respect to bulk water, neither in the first nor in the second
hydration shell. 
If the main effect of a hydrophobic solute would
be the reduction of the configuration space
  for hydrogen bonding, then one would naturally expect to find less
  neighbors in close contact. Clearly, this is not the case. 
Moreover, the 
distribution is much narrower in bulk water,
which already suggests that the small-scale fluctuations in the
vicinity of hydrophobes
(i.e. librations, hydrogen-bond exchange, etc.) are enhanced with
respect to the bulk. According to our criterion
for the nearest neighbor we find that the probability of a water molecule 
having 3 hydrogen-bonded nearest neighbors is negligible, irrespective
of its position.       

Aside from an altered number of nearest neighbors the traditional
  picture also suggest a more ordered local structure. In order to
  cause entropy loss the structural fluctuations should
  tend to diminish. To inspect in detail the structural
ordering of water 
molecules in close contact, around hydrophobic particles and in
bulk water, we employ the recently introduced dipolar order
parameter \cite{PRL}, 
\begin{equation}
D(i) = \frac{1}{N}\sum_{j=1}^N\left(
\frac{\alpha_{ij}-\alpha_{ij}^{min}}{\alpha_{ij}^{max}-\alpha_{ij}^{min}}
\right) 
\label{eq1}
\end{equation}
where $i\ne j$ and the sum is taken over the $N$ neighbors in the first
coordination shell of the $i$-th water molecule and
$\alpha_{ij}^{max}$ and
$\alpha_{ij}^{min}$ correspond to maximal and 
minimal dipole-dipole potential at a given intermolecular unit
vector and unit dipole vector of the tagged molecule. 
This way, $D(i)$ takes values between 0 (maximal
repulsion) and 1 (maximal attraction).
The distributions of dipolar ordering shown in Fig. \ref{fg2}
confirm the idea that water is
on average slightly more orientationally ordered in the first
hydration shell of hydrophobic
particles (note the slight shift of the distributions towards higher $D$
values). More importantly, the width of the distributions $p(D)$, which
reflects the constraining of orientational degrees of freedom,
remains rather unaffected. This
width clearly increases upon increasing polarity of the solute
indicating an orientational relaxation in the hydrogen-bond network
\cite{PRL}. 
The situation in
the second shell is similar, albeit less pronounced. Most likely, due 
to a commensurability effect, the second shell molecules exhibit 
a reverse trend in orientational ordering in the case of $s2$. 

Apparently 
the pair interactions and consequently also the hydrogen bonds are
strengthened but there is no 
reduction of the orientational configuration space explored by
individual water molecules in the hydration shells as the
distributions are merely shifted while their form remains unchanged. 
Thus, we find that {\it i)} there are no significant differences in
the number of nearest HB and non-HB neighbors and {\it ii)} there 
is no orientational constraining although the local structure is 
more ordered. This is clearly
in contradiction with the idea of reduced configuration space
available for hydrogen bonding \cite{Still,Chand}. So how can
this contradiction be reconciled?

There is an important difference between the total volume of 
configuration space and the volume that is actually 
visited at a given temperature.
As the number of close contacts is not altered significantly 
by the mere presence of the
solute the latter is not expected to significantly affect the volume
of the configuration space
available to a water molecule for hydrogen bonding.  
On the other hand, in the liquid state water
molecules form a dynamic, 
labile hydrogen-bond network. 
There is clear evidence that the molecular reorientation underlying
the exchange of hydrogen bonds is not diffusive ({\it i.e.} does not
occur as angular Brownian motion) but rather proceeds in terms of
large-amplitude sudden jumps
\cite{Laage1,Laage2,Eaves,Laage3,Moilanen,Ji,Laage4}. 
As it is impossible for water molecules to each form four
HBs at a time in the liquid state, this introduces a strong
frustration into the system 
emerging from the competition between the
entropic and energetic driving forces. 
This means, that while fluctuating about the minimum energy
configuration exhibiting {\it e.g.} librational movements, a 
water molecule is most likely to form two hydrogen-bonds to 
its closest neighbors.  As a water molecule in the bulk liquid 
is intrinsically inclined to form more than one hydrogen bond,
this necessarily leads to confinement of its orienational 
configuration space. This fact
  suggests to rationalize the entropy loss in HE in an
  alternative way.

The current understanding of hydrogen-bond network dynamics in
neat water already anticipates the collective concerted motions
of several water molecules \cite{Laage4}. This suggests looking for
the potential source of
entropy loss in the altered 'communication' between emerging
transient water clusters involved in the exchange of hydrogen
bonds, {\it i.e.} the reduction of configuration space due to
constraining of collective degrees of freedom. For large enough
fluctuations the transient water cluster can 
rearrange into another configuration just by the exchange of
hydrogen bonds. Hydrogen-bond fluctuations (for example, librations)
try to compensate for the entropic frustrations caused by a
transient mutual alignment of water molecules in between sudden
jumps. Since structural fluctuations are 
caused by local field
fluctuations they are expected to be dramatically suppressed in direct
vicinity of a hydrophobic solute, because the latter exerts only
an extremely weak electrostatic field. Without an additional
compensation mechanism the hydrogen bonds would be significantly
strengthened and the entropic frustrations would be expected to grow
further. Therefore there must exist a tendency of nearby water
molecules (nearest and next-nearest neighbors) to compensate for the
suppressed fluctuations of local electrostatic fields. This in turn can
not happen unless the fluctuations of nearby hydrogen-bonded clusters
(intra- and intercluster fluctuations)
become correlated as to maximize local field fluctuations while
maintaining as many as possible mutual water arrangements close to the
optimal HB geometry. Thus, in order to satisfy the {\it local energetic
  demand} to form hydrogen bonds, the resulting {\it entropic
  frustration relaxes non-locally} in the vicinity of a hydrophobic
solute.             

Testing this hypothesis demands the evaluation of various
many-body correlations. Although addressing the problem of coupled
translational-orientational multi-body correlations is nontrivial 
it can be significantly simplified in the following manner. 
While single water molecules are, by nature,
indistinguishable they can be {\it transiently} classified as being
hydrogen bonded or non-bonded to its nearest neighbor (using the same
criteria as above). Any HB exchange event merely permutes the indices
between water molecules.
With these criteria we can, at any instant,
classify the HB and non-HB neighbors of any given molecule. Involving
various types of fluctuations on different time-scales, such as
librations, dimer tumbling and HB jumping (see for example
\cite{Laage4} and references therein), following both the relative
positions and orientations of water molecules individually is simply too
complex and in fact unnecessary. To determine how molecular
degrees of freedom are constrained in the liquid state we shall be
primarily interested in structural fluctuations in water clusters.
Such fluctuations can be easily described in terms of
fluctuations of interaction energies of hydrogen bonded and
non-hydrogen bonded molecules.

Using the
random variable transformation theorem \cite{RVT} we can map the
joint configurational probability density onto its functionally
dependent joint probability 
density for pair potential energies. Thereby a given matrix of
molecular positions and orientations, $\boldsymbol{\Omega}$, is mapped
onto a vector of interaction energies, $\bm{U}$, which can be formally
written for given values $\bm{\omega}$ and $\bm{u}$ as follows:
\begin{equation}
p(\bm{u}) = \int \mathrm{d}\boldsymbol{\omega}^N
P(\boldsymbol{\omega}^N)\delta(\bm{u}-\bm{f}(\boldsymbol{\omega}^N)),  
\label{eq2}
\end{equation}     
The functional relation $\delta(\bm{u}-\bm{f}(\boldsymbol{\omega}^N))$ contains
a class indicator (close contact, hydrogen-bonded, etc.) as well as
the appropriate averaging operation accounting for the
indistinguishability of pairs within a certain class (for details see
section 1 in the Supplementary information). This way we can
construct, using appropriate 
functional relations, probability densities for observing, for
example, a (non)hydrogen bonded pair with certain energy or a joint
probability density of observing two pairs of different classes having
given energies. 
Comparing the values per
water molecule/cluster inside the first and second hydration shell with the
corresponding bulk values we are able to select the most important
contributions to the entropy loss. The Gibbs-Shannon entropy, 
$S[p]=-k_B\int p(u_i)\ln p(u_i)du_i$, is used to evaluate the total
uncertainty of a quantity $u_i$, and thus quantifies the fluctuations. 
Alternatively, Kullback-Liebler entropy or {\em
  Correlation entropy} (the latter is taken after \cite{Gu}) is used
to quantify the total correlation between two random variables $X$ and
$Y$ (which can be components of $\bm{U}$, for example various
combinations of hydrogen-bonded and non-HB pairs ($U_{HB}$ and
$U_{non-HB}$). The correlation entropy can be expressed in terms of
individual and joint entropies, $S_C(X,Y)=S(X)+S(Y)-S(X,Y)$, or
explicitly in terms of corresponding probability densities: 
\begin{equation}
S_c(X,Y) = k_B \iint p(x,y)\ln
\frac{p(x,y)}{p(x)p(y)} \mathrm{d}x\mathrm{d}y.  
\label{eq3}
\end{equation}
We limit the present discussion to 3 and 4-particle correlations
to asses how fluctuations of various HB and non-HB pairs are
correlated.

\begin{figure}
\includegraphics[width=0.98\textwidth]{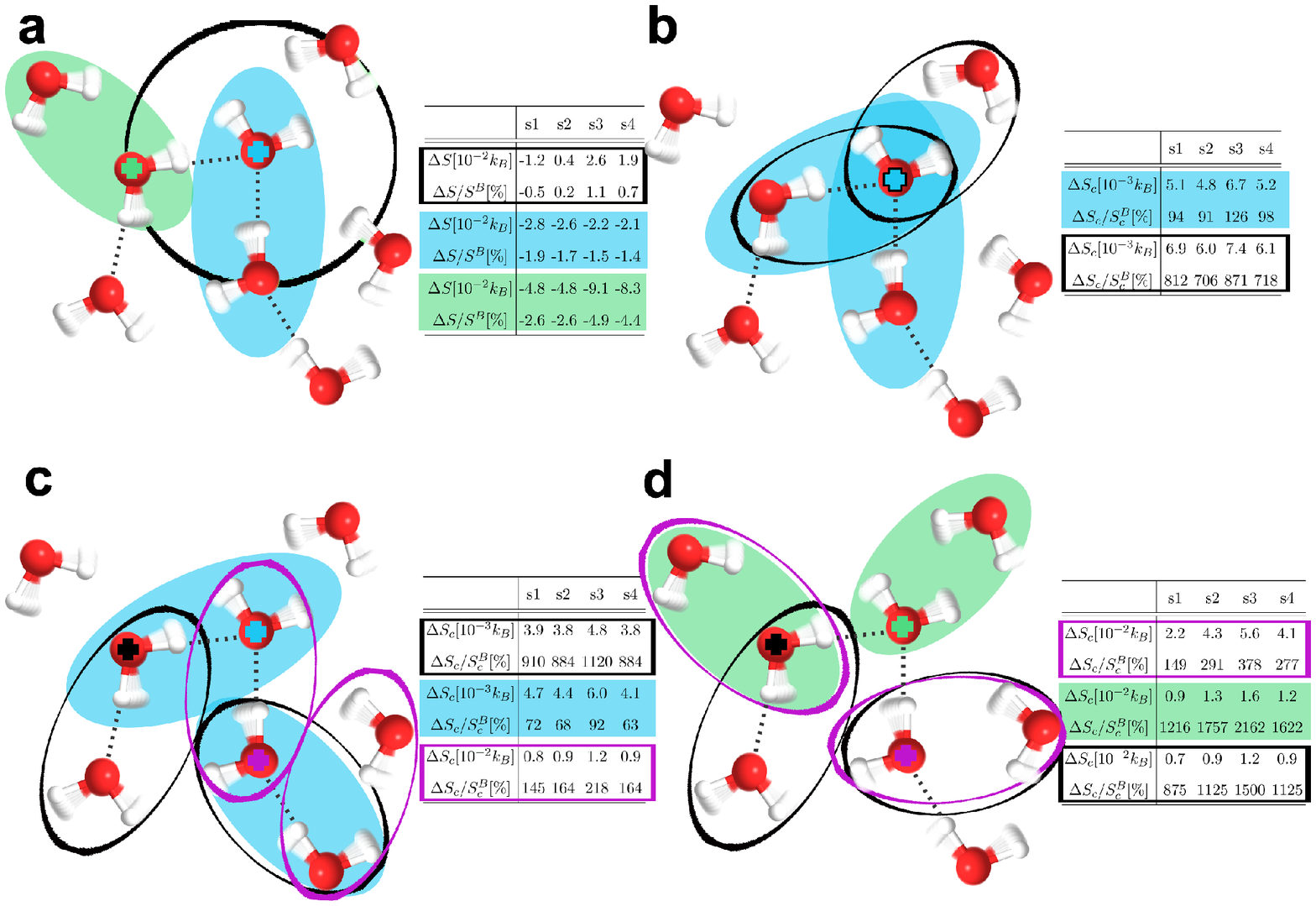}
\caption{}
\label{fg3}
\end{figure}

Being predominantly interested in generic features and less on the specific
effect of solute sizes we focus first on the properties of individual
pairs of molecules and 
find that the dipolar entropy difference between the hydration shells and the
bulk, which directly measures the degree of orientational constraining, 
does not show a monotonic behavior with respect to the solute
size and is less than 1\% of the corresponding bulk value (for details
see section 2 in the Supplementary information). This
confirms the assumptions that a reduced orientational configuration
space of single molecules is {\it not} the reason for HE. This fact is further
substantiated with the values of the entropy difference, $\Delta
S_{tot-c}$, of the
fluctuations of the total interaction energy of tagged molecules
with its nearest neighbors, $S[p(\sum_iU_{i0})]$, (Fig
\ref{fg3}a). It may also be taken 
as a contribution to the fluctuations of the local electrostatic field
(solely) due
to the thermal motion of nearest neighbors. In fact, $\Delta
S_{tot-c}$ is mostly positive, except for the 
smallest solute which indicates that local field fluctuations due to
nearest neighbors are enhanced in the vicinity of the
hydrophobe. Meanwhile,  
the entropy of mutual fluctuations of the potential energy of both
hydrogen bonded neighbor pairs (Fig. \ref{fg3}a blue) and  non-hydrogen
bonded pairs (Fig. \ref{fg3}a green) decrease with respect to bulk
water. The former can be understood as a measure of
librational-type of fluctuations and its lowering is indicative of a
slight HB
strengthening \cite{Moilanen} (though the relative difference is
smaller than 2$\%$ of the bulk value). Together with decreasing fluctuations
of the non-HB interaction energy and the simultaneous increase of
fluctuations of the total interaction energy with nearest
neighbors this immediately hints at higher order (beyond pair)
correlations, as suggested by our hypothesis.  
The
local strengthening of HB is also suggested by the shift of $p(D)$
towards larger $D$. The fact that fluctuations of the total
interaction energy are so small and their entropy difference
mostly even positive along with the strengthening of HB are fully
consistent with the large-amplitude sudden jump picture of HB-network
dynamics.    
A closer look at entropies of various 3- and 4-body correlations in
Fig \ref{fg3}b-d  reveals 
striking, upto 22 fold increases with respect to bulk water. The largest
increase is observed in the case of correlations of nearby non-HB
pairs bridged via a HB (12-22 fold), a HB pair and a nearby non-HB
pair (7-9 fold 9-15 in the case of geminal and  9-15 fold in the case
of vicinal pairs) and two HB pairs that do not share a common HB bond
(are not bridged via a HB; 9-11 fold). A significant, albeit smaller,
increase is also 
observed in the case of geminal HB pairs (app. 100$\%$ increase) and HB
pairs which are bridged via a HB (70-90$\%$ increase). Let us briefly
mention that all 3- and 4-body correlations are, to
linear order, anti-correlated.
The pronounced many-body correlations
propagate into the second hydration shell (for details see Table 1 in section 3
of the
Supplementary information) but fall off rapidly further away. Thus, the
picture of two perturbed hydration layers is retained. 

\begin{figure}
\includegraphics[width=0.98\textwidth]{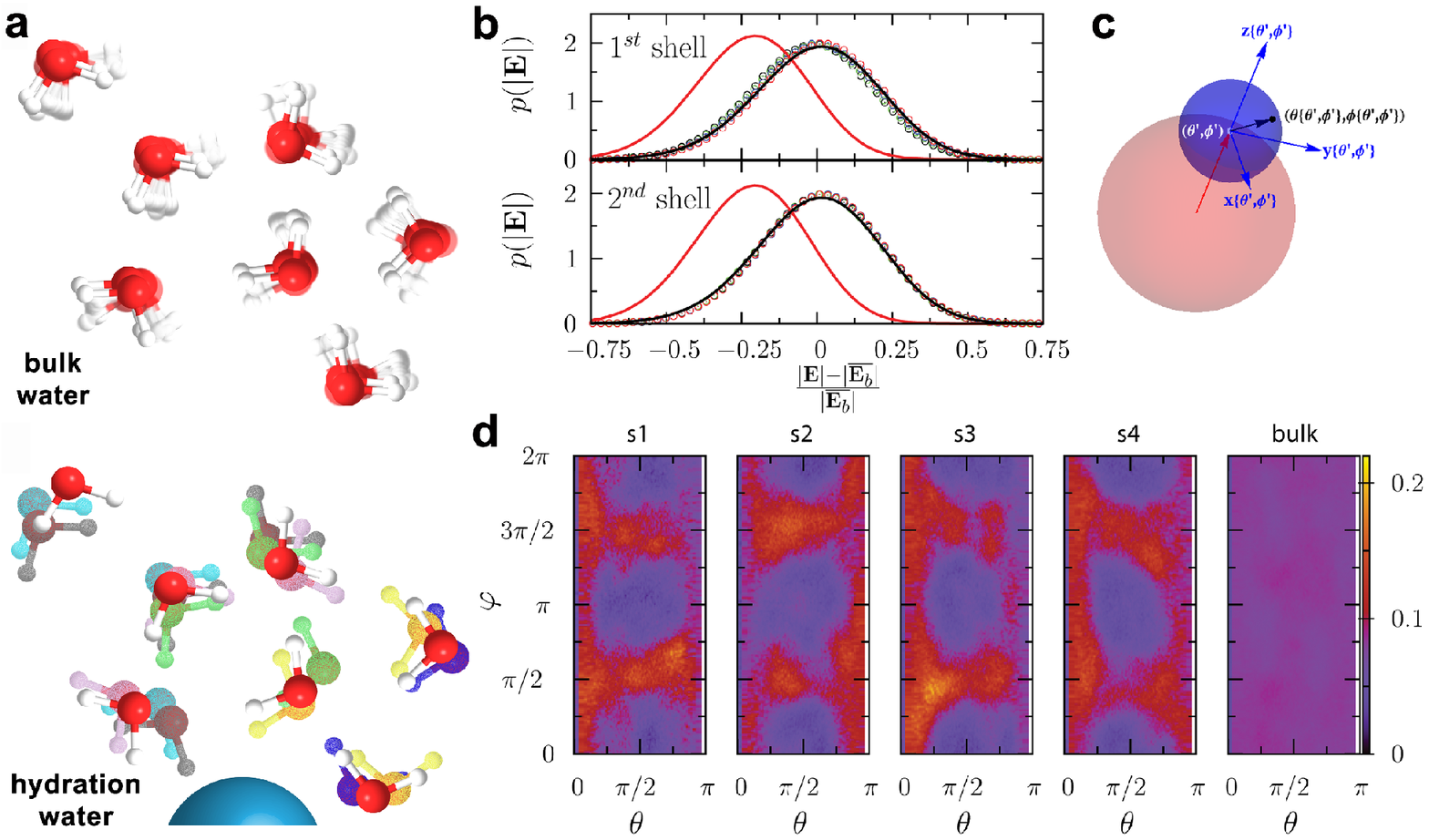}
\caption{Fig 4}
\label{fg4}
\end{figure}

The results suggest that the traditional explanation of the HE on the
molecular scale needs a substantial revision. It is not a reduced
configuration space for hydrogen bonding that is responsible for the
observed lowering of entropy, but a striking increase of many-body
correlations (Fig. \ref{fg4}a), 
which is essentially due to hydrogen bonding being a strong and orientationally
dependent interaction. The increase of many-body correlations
is necessary to compensate for the reduction of
fluctuations in the local electrostatic field and the resulting local
HB-strengthening when one or more ''polar'' water
molecules are replaced by an ''apolar'' hydrophobic
particle. This effect should scale as the difference between local
field distributions experienced by the water molecule in bulk
(Fig. \ref{fg4}b black curve) and by
a bulk water molecule for which one neighboring water molecule is
omitted from the calculation of the local field (Fig. \ref{fg4}b black curve). If left
uncompensated, such as in the latter case, the distribution is
shifted significantly to lower values. It also 
turns out that the distributions of field strengths in the hydration
shells are almost identical to the one in bulk liquid (Fig. \ref{fg4}b symbols).
Furthermore, by looking at the 
distribution of orientations of the local field over the unit sphere
(for details see Fig. \ref{fg4}c and section 2 of the Supplementary
information) shown in Figure \ref{fg4}d we find that the distribution
is isotropic in the bulk as well as in the 
second hydration shell (not shown).
Meanwhile, in the first hydration shell the local field
fluctuations are
constrained to the plane containing the normal to the solute
surface. Such a distribution ensures that all water
molecules in the first hydration shell experience the maximum 
span of local fields 
and thereby local torques which
counteract the entropic frustrations caused by a
transient mutual alignment of water molecules in between HB exchange events. 
Such a compensation mechanism comes at cost of 
many-body correlations.
The local increased entropic frustration is thus compensated
non-locally.       

The proposed picture of hydrophobic hydration
does not imply any water
immobilization in the spirit of the iceberg hypothesis. 
In explaining the moderate
retardation of reorientation dynamics \cite{Laage3,Halle}
the current picture, based on the excluded volume effect, 
proposes that the slowdown is a result of fewer
accessible configurations of the transition-state (TS) in a HB exchange
event due to the presence of the solute. We demonstrate that 
the lower
number of accessible TS configurations rather results from many-body
correlations ({\it due to correlations the TS is simply statistically
less probable}) rather than actual sterical hindrance of
reorientation. Thus the expression for the slowdown factor remains
unaffected. Our view also provides an explanation of
the slightly
better solubility of nonpolar solutes in heavy relative to
light water \cite{Graziano}. While being more inert heavy water
responds less strongly to local field fluctuations, which leads to HB
strengthening and thus a higher cohesive energy density \cite{Graziano}
but also results in a more structured liquid with respect to light
water \cite{Soper}. Being intrinsically more structured the relative
extent of local field fluctuations to be compensated upon the
introduction of a hydrophobic molecule is therefore also smaller.

Finally, the results allow us to speculate that the range of
hydrophobic interactions beyond that expected from the
minimally-exposed surface area reasoning is a result of the
propagation of many-body correlations beyond the first hydration shell.

\section{Methods}

We performed extensive constant pressure Monte-Carlo (MC) simulations with TIP5P
water and freely moving hard sphere solutes at room temperature (298
K) and 1 atm assuming periodic boundary conditions.
After an extensive equilibration period we performed as many trial moves as
to assure that on average each molecule was {\it successfully} moved $4
\times 10^5$ times at the maintained acceptance rate of 30$\%$. To
exclude (auto)correlations  the
successive configurations used for the analysis were taken to be as
far apart as to assure that between each taken configuration each molecule
was {\it successfully} moved at least 5 times. The hard sphere radius
of the water molecule when in close contact with the hydrophobic
spheres was taken to be 1.4 \AA . The following radii of solutes were
considered: r(s1)=1.4 \AA , r(s2)=2.1 ($1.5 \times r_{H_2 O}^{hs}$)
\AA , r(s3)=2.52 
($1.8 \times r_{H_2 O}^{hs}$) \AA and r(s4)=2.8 ($2 \times r_{H_2 O}^{hs}$)
\AA . According to the solute size the number of water molecules was
(in ascending size) 2668, 2670, 3513 and 3560. The total number of MC
steps was $35.57 \times 10^8$, $35.60 \times 10^8$, $46.8 \times 10^8$
and  $47.5 \times 10^8$. 

\nocite{*}

\bibliography{apssamp}

\section{Acknowledgements}
We thank U. Maver for his help in preparation of schematics
and professors M. R. Johnson, M. Gaber\v s\v cek and P. Ziherl for 
suggestions and critical reading of the manuscript.

\clearpage
\newpage

Figure 1: {\bf Radial ordering around hydrophobic particles.} {\bf a}
  Radial correlation function for water centers; (inset) corresponding
  relative density fluctuations. (bottom) Distribution of the number
  of total (dashed lines) and hydrogen-bonded (full lines) contacts
  per water molecule located in the {\bf b} first and {\bf c} second
  hydration shell with the corresponding bulk water values for
  comparison. The vertical lines denote expected values. The solute
  sizes are: r(s1)=1.4 \AA , r(s2)=2.1 ($=1.5 \times r_{H_2 O}^{hs}$)
  \AA , r(s3)=2.52 ($=1.8 \times r_{H_2 O}^{hs}$) \AA and r(s4)=2.8 ($=2 \times r_{H_2
    O}^{hs}$) \AA (see Methods for details).\\
\\

Figure 2: {\bf Mutual orientational ordering of water molecules in
    close contact.} Difference in the distribution of the
  dipolar order parameter with respect to the bulk, $\Delta
  p(D)=p(D)-p_B(D)$,
in the first {\bf a} and second {\bf b}
  hydration shell. The dipolar order distribution in bulk water is
  shown in the inset of {\bf a}. {\bf c} Joint probability of finding
  two molecules at a distance $r$ apart having a (pair) mutual order
  $D_2$ (see Eq. \ref{eq1} for definition, with $N$=1).\\
\\

Figure 3: {\bf Fluctuation and correlation entropy differences per
    tagged pair/cluster in the first hydration shell with respect to
    the bulk are given in the Tables. Correlations between differently
interacting pairs of water molecules are depicted by colour in
schematics and Tables.} Only the tagged
  molecule necessarily lies in the first hydration shell. The crosses
 denote the tagged molecule which is used for the
localization of a given cluster and the dashed black lines denote
HBs. {\bf a} Absolute and relative entropy 
difference of fluctuations of HB (blue shading), non-HB (green
shading) and total 
interaction energy with nearest neighbors (black frame). The slight increase
of the entropy of total interaction energy fluctuations accompanied
with the simultaneous decrease of entropy of HB and non-HB energy
fluctuations is indicative of higher order correlations. {\bf b}
3-body correlation entropy of geminal (pairs share the tagged
molecule) HB pairs (blue shading) and a 
geminal HB and non-HB pair (black frame). {\bf c} 4-body correlation entropy
of HB-bridged HB pairs (blue shading), non-HB-bridged HB pairs (black
frame) and a vicinal (pairs do not share the tagged molecule) tagged HB and
non-HB pair (magenta frame). {\bf d} 4-body correlation entropy 
of vicinal HB-bridged non-HB pairs (green shading), vicinal
non-HB-bridged non-HB pairs (magenta frame) and a vicinal tagged
non-Hb and HB pair (black frame). The upto 2200\%\ increase of many-body
correlations with respect to bulk water confirms their crucial role in
mediating the hydrophobic effect.\\
\\

Figure 4: {\bf Emergence of pronounced many-body correlations and their
    physical origin.} {\bf a} In
  bulk water the fluctuations are correlated predominantly up to the
  pair level. In the vicinity of hydrophobic solutes
  significant many-body correlations emerge in order to compensate for
  reduced 
fluctuations in the local electrostatic field as a result of the
exceedingly small field of the hydrophobic solute. Different colours
    denote various orders of correlated 
fluctuations. {\bf b} Distribution of magnitudes of the local
electrostatic field experienced by a water molecule at a given
location. Black lines denote distributions in bulk water and symbols
represent results for water molecules around apolar solute.
If one nearest neighbor water
molecule is omitted from the calculation of the local field
in the bulk, this leads to uncompensated fluctuations shown by the red
distribution. Results are
given as relative deviations from the average bulk value
$|\overline{{\bm E}_b}|$. Note that the strength of the local field
experienced by water molecules in the first and second hydration shell
is almost identical to the one in bulk water. {\bf c} The coordinate
frame used in the 
calculation of angular distributions of local electrostatic
fields. The radial projection 
of the water molecule position onto a
spherical surface $(\theta',\varphi')$ defines
the origin and orientation of a secondary coordinate
system (blue), such that the secondary ${\bm z}(\theta',\varphi')$,
${\bm x}(\theta',\varphi')$ and ${\bm y}(\theta',\varphi')$  axes correspond to
the local radial, polar and azimuthal directions. Angular coordinates
in the secondary frame 
$(\theta \{\theta',\varphi'\},\varphi \{\theta',\varphi'\})$ are
used to describe the distribution of the local field over the surface
of the unit sphere. {\bf d} Distribution of orientations of the local
electrostatic field in the first
hydration shell of apolar solutes and in bulk water. In the first
hydration shell the orientational fluctuations of the local field are
confined to the plane containing the normal to the solute surface. The
distribution in the bulk as well as in the second hydration shell (not
shown) is entirely isotropic.

\clearpage
\newpage

\end{document}



\title{Supplementary information for: Physical origin underlying the entropy loss upon hydrophobic hydration}

\author{Alja"z Godec}
\email{aljaz.godec@ki.si}
\author{Franci Merzel}
\email{franci.merzel@ki.si}
\affiliation{National Institute of Chemistry, Hajdrihova 19,1000 Ljubljana, Slovenia}



\date{\today}
\maketitle

\section{Probability densities and the mapping onto potential energy space}
Throughout the paper as well as this section we strictly discriminate between a random
variable, which is denoted by uppercase letters, and a value it can
take on (lowercase letters). The
random variable transformation theorem is used to map the
joint configurational probability density, onto its functionally dependent joint probability
density for pair potential energies. Thereby molecular positions and
orientations, $\boldsymbol{\Omega}^N=(\bm{R}^N,\boldsymbol{\alpha}^N)$,
are mapped
onto a vector of interaction energies, $\bm{U}$:
\begin{equation}
p(\bm{u}) = \int \mathrm{d}\boldsymbol{\omega}^N P(\boldsymbol{\omega}^N)\delta(\bm{u}-\bm{f}(\boldsymbol{\omega}^N)),  
\label{eq1}
\end{equation}     
The functional relation $\delta(\bm{u}-\bm{f}(\boldsymbol{\omega}^N))$ contains
class indicators (close contact, hydrogen-bonded, etc.) and the
the appropriate averaging operation accounting for the
indistinguishability of pairs within a certain class. 

If we denote the solute position as $\bm{R}_s$ the general 
functional expressions $\delta(\bm{u}-\bm{f}(\boldsymbol{\omega}^N))$
can be written as follows.

In the scalar case ({\it i.e.} density $p(U_x)$) we have:

\begin{equation}
\delta(u-f(\boldsymbol{\omega}^N))\equiv
\frac{   \sum_{i,j\ne
  i}\delta(u-u(\boldsymbol{\omega}_i,\boldsymbol{\omega}_j))\chi_d(\boldsymbol{\omega}_i,\boldsymbol{\omega}_j)
\chi_{HB}(\boldsymbol{\omega}_i,\boldsymbol{\omega}_j)\Pi_{R_{min},R_{max}}(|\bm{r}_i-\bm{r}_s|) }
{\int \mathrm{d}\boldsymbol{\omega}^N P(\boldsymbol{\omega}^N)\left \{ \sum_{i,j\ne
  i}\chi_d(\boldsymbol{\omega}_i,\boldsymbol{\omega}_j)
\chi_{HB}(\boldsymbol{\omega}_i,\boldsymbol{\omega}_j)\Pi_{R_{min},R_{max}}(|\bm{r}_i-\bm{r}_s|)\right
\} }
\label{eq2}
\end{equation} 
where $\chi_d(\boldsymbol{\Omega}_i,\boldsymbol{\Omega}_j)$ is the
indicator function for the distance cutoff for nearest neighbors
\begin{equation}
\chi_d(\boldsymbol{\omega}_i,\boldsymbol{\omega}_j)\equiv (1-H(|\bm{r}_i-\bm{r}_j|)),  
\label{eq3}
\end{equation}
$H(x)$  being the Heaviside function and
$\chi_{HB}(\boldsymbol{\omega}_i,\boldsymbol{\omega}_j)$
is the indicator function for hydrogen bonding (which is left out in
the case of non-hydrogen bonded neighbors)
\begin{equation}
\chi_{HB}(\boldsymbol{\omega}_i,\boldsymbol{\omega}_j)\equiv H(\vartheta[OH_i\cdots O_j]-\vartheta_{min})+H(\vartheta[OH_j\cdots O_i]-\vartheta_{min}),  
\label{eq4}
\end{equation}
where $\vartheta[OH_i\cdots O_j]$ is the angle between the OH bond of
the donor molecule and vector connecting the proton of the donor
molecule and the acceptor oxygen. $\vartheta_{min}$ is the lower cutoff for the hydrogen bonding
angle. $\Pi_{a,b}(x)$ represents the boxcar function
\begin{equation}
\Pi_{R_{min},R_{max}}(|\bm{r}_i-\bm{r}_s|)\equiv \begin{cases}
0 & \text{if } |\bm{r}_i-\bm{r}_s| < R_{min},\\
1 & \text{if } R_{min}\le |\bm{r}_i-\bm{r}_s|\le R_{max},\\
0 & \text{if } |\bm{r}_i-\bm{r}_s| > R_{max}.
\end{cases} 
\label{eq5}
\end{equation}
which localizes the tagged molecule $i$ in a certain hydration shell
(first, second shell or bulk). 

In the non-scalar case
$p(\bm{U})$ is a joint distribution. Then we have, for example
for a joint distribution for a 3-body or 4-body density (or
equivalently for a pair density of pairs), $p(U_{0x},U_{0y})$ or
$p(U_{0x},U_{jz})$ respectively:
\begin{equation}
\delta(\bm{u}-\bm{f}(\boldsymbol{\omega}^N)) \equiv 
\frac{ \sum_{\substack{ijkl}}\delta(u-u^{ij})\delta(u-u^{kl})
\chi_d^{ij}\chi_d^{kl}\chi_{HB}^{ij}\chi_{HB}^{kl}\Pi_{R_{min},R_{max}}(d_i)(1-\delta_{il}\delta_{jl}\delta'_{ik}\delta_{jk})\Theta_c^{ijkl}}
{\int \mathrm{d}\boldsymbol{\omega}^N P(\boldsymbol{\omega}^N)\left \{
  \sum_{\substack{ijkl}} \chi_d^{ij}\chi_d^{kl}\chi_{HB}^{ij}\chi_{HB}^{kl}\Pi_{R_{min},R_{max}}(d_i)(1-\delta_{il}\delta_{jl}\delta'_{ik}\delta_{jk})\Theta_c^{ijkl}
  \right \}}, 
\label{eq6}
\end{equation}
where the explicit coordinate dependences
$(\boldsymbol{\Omega}_i,\boldsymbol{\Omega}_j)$ were replaced by
superscripts $^{ij}$, $d_i=|\bm{r}_i-\bm{r}_s|$ and $\delta_{ij}$ is
the Kronecker delta. The primed Kronecker delta denotes that $i$ may
or may not be equal to $k$, depending on whether we take geminal or
vicinal pairs of molecules, that is, 3- or 4- particle correlations
(for the explicit definition of geminal and vicinal pairs see main
text). For geminal pairs the primed Kronecker delta is omitted. $\Theta_c^{ijkl}$ is the connectivity indicator function and
defines the topological relation between both pairs. Two pairs may be 
connected with a hydrogen bond or not.

Corresponding equation for higher dimensional probability densities
can be constructed accordingly.

\section{Calculation of the orinentational distribution of the local
  electrostatic field}

The systems under investigation all have spherical symmetry around the
center of the solute. Thus, all points in the space-fixed frame
(denoted by prime) with
with the same radius $r'$ are positionally equivalent. As we classify
the water molecules according to their position with respect to the
solute as belonging to the first and second hydration shell or the
bulk, we can drop the radial component when dealing with molecules in
a particular shell and simply retain the angular variables
$(\theta',\varphi')$, where $\theta'=\arccos(\frac{z'}{r'})$ and
$\varphi'=\arctan(\frac{y'}{x'})$. We choose to describe the
orientational distribution of the instantaneous
electrostatic filed orientation in a given point of the space-fixed
frame as a distribution over the surface of a unit sphere. In order to
take into account the equivalence of positions at given radius in the
space-fixed frame we set the secondary local frame in the following
manner. The secondary $x,y$ and $z$ axes are chosen to coincide with
the polar, azimuthal and radial directions at $(\theta',\varphi')$. Specifically, this amounts to the following basis:
\begin{equation*}
\mathbf{\hat{x}}(\theta',\varphi') = \left(
\begin{array}{c}
\cos \varphi' \sin \theta'\\
\sin \varphi' \cos \theta'\\
-\sin \theta'
\end{array} \right),
\qquad
\mathbf{\hat{y}}(\theta',\varphi') = \left(
\begin{array}{c}
-\sin \varphi'\\
\cos \varphi'\\
0
\end{array} \right),
\qquad
\mathbf{\hat{z}}(\theta',\varphi') = \left(
\begin{array}{c}
\cos \varphi' \sin \theta'\\
\sin \varphi' \sin \theta'\\
\cos \theta'
\end{array} \right).
\end{equation*}
The local electrostatic field at a point ${\bm r'}=(x',y',z')$ due to
a collection of $N$ point charges $q_i$ with coordinates $\bm r''_i$ is
${\bm E}(\bm r')=\sum_{i=1}^N\frac{q_i({\bm r}'-{\bm r}''_i)}{|{\bm
    r}'-{\bm r}''_i|^3}$. The unit Cartesian components of the field in the secondary
frame at $(r',\theta',\varphi')$ are thus: 
\begin{equation*}
\hat{E_x}(r',\theta',\varphi')=\frac{ ({\bm  E}(\bm r')\cdot \mathbf{\hat{x}})}{|{\bm  E}(\bm r')|},
\qquad
\hat{E_y}(r',\theta',\varphi')=\frac{ ({\bm  E}(\bm r')\cdot \mathbf{\hat{y}})}{|{\bm  E}(\bm r')|},
\qquad
\hat{E_z}(r',\theta',\varphi')=\frac{ ({\bm  E}(\bm r')\cdot \mathbf{\hat{z}})}{|{\bm  E}(\bm r')|}.
\end{equation*}
After a trivial transformation to spherical coordinates in the
secondary frame, we may write for the orientational distribution of
the electrostatic field experienced by a water molecule in shell $d_i$
(first, second or bulk):
\begin{equation*}
p(\theta,\varphi)[d_j]=\left \langle
\frac{\sum_{i=1}^{N}\delta (\hat{E_{\theta,i}}-\theta)\delta (\hat{E_{\varphi,i}}-\varphi)\Pi_{R_{min},R_{max}}(d_i)}{\sum_{i=1}^{N}\Pi_{R_{min},R_{max}}(d_i)}
\right \rangle .
\end{equation*}

\section{Gibbs-Shanon and correlation entropy differences}
We denote a tagged molecule with $0$, its neighbors
with $i,j,k$; $i\ne j \ne k \ne \ldots$, and the nearest neighbors of
the nearest neighbors with primes.
A 3-body correlation entropy of a tagged molecule and its two nearest
HB neighbors is thus, for example, $S_{HB-HB}^{i0-0j}$, and the 4-body correlation
entropy of two adjacent HB pairs is $S_{HB-HB}^{i0-jk'}$, if $j$ is not
hydrogen-bonded to $0$, and $S_{HB-HB}^{i0-jk'*}$, if $j$ is a HB
neighbor of $0$.
.

\begin{table*}[ht]
\caption{Entropy differences per water molecule in the first and
  second shell for various solute sizes. $\Delta S_i$ stands for
  $S^{i}-S_B^{i}$ and the asterisk denotes that molecule $j$ is
  hydrogen-bonded to the tagged molecule $0$.}
\begin{center}
\begin{tabular}{|c||c|c||c|c||c|c||c|c|}
\hline
 & s1 $1^{st}$& s1 $2^{nd}$& s2 $1^{st}$& s2 $2^{nd}$& s3
  $1^{st}$& s3 $2^{nd}$& s4 $1^{st}$ & s4 $2^{nd}$ \\
\hline
\hline
$\Delta S_{D} [10^{-3}k_B]$  & -7.9  & -6.4 & 4.5 & 2.0 & -4.6  &
-6.1 & 5.6  & 3.3 \\
$\Delta S_{D}/S_D^B [10^{-1} \%]$ & 8.6  & 6.9 & -4.9 & -2.2 & 5.0 & 6.7  & -6.2  & -3.6 \\
\hline
$\Delta S_{tot-c} [10^{-2}k_B]$ & -1.2 & -0.3 & 0.4 & 0.2 & 2.6 & -0.3 & 1.9 & 0.2\\
$\Delta S_{tot-c}/S_{tot-c}^B [\%]$ & -0.5 & -0.1 & 0.2 & 0.1 & 1.1 & -0.1 & 0.7 & 0.1 \\
\hline
$\Delta S_{HB} [10^{-2}k_B]$ & -2.8 & -1.1 & -2.6 & 0.4 & -2.2 & 0.07 & -2.1 & -0.5\\
$\Delta S_{HB}/S_{HB}^B [\%]$ & -1.9 & -0.7 & -1.7 & 0.3 & -1.5 & 0.05 & -1.4 & 0.3 \\
\hline
$\Delta S_{nHB} [10^{-2}k_B]$ & -4.8 & -0.5 & -4.8 & -0.1 & -9.1 & -0.3 & -8.3 & -0.9\\
$\Delta S_{nHB}/S_{HB}^B [\%]$ & -2.6 & -0.3 & -2.6 & -0.05 & -4.9 & -0.16 & -4.4 & -0.5 \\
\hline
$\Delta S_{HB-HB}^{i0-0j} [10^{-3}k_B]$ & 5.1 & 1.7 & 4.8 & 1.3 & 6.7 & 1.9 & 5.2 & 1.8\\

$\Delta S_{HB-HB}^{i0-0j}/S_{HB-HB}^{i0-0j;B} [\%]$ & 94  & 31 & 91 & 25
& 126 & 36 & 98 & 34\\
\hline
$\Delta S_{HB-nHB}^{i0-0j} [10^{-3}k_B]$ & 6.9 & 2.0 & 6.0 & 1.8 & 7.4 & 2.2
& 6.1  & 1.7\\

$\Delta S_{HB-nHB}^{i0-0j}/S_{HB-nHB}^{i0-0j;B} [\%]$ & 812 & 235 & 706 &
212 & 871 & 256 & 718 & 200\\ 
\hline
$\Delta S_{HB-HB}^{i0-jk'} [10^{-3}k_B]$ & 3.9 & 1.2 & 3.8 & 1.1 & 4.8
& 1.3 & 3.8 & 1.1 \\

$\Delta S_{HB-HB}^{i0-jk'}/S_{HB-HB}^{i0-jk;B} [\%]$ & 910 & 440 & 884
& 260 & 1120 & 320 & 884 & 256 \\
\hline
$\Delta S_{HB-HB}^{i0-jk'*} [10^{-3}k_B]$ & 4.7 & 1.2 & 4.4 & 0.9 & 6.0
& 1.4 & 4.1 & 1.4 \\

$\Delta S_{HB-HB}^{i0-jk'*}/S_{HB-HB}^{i0-jk*;B} [\%]$ & 72 & 18 & 68
& 14 & 92 & 22 & 63 & 22\\ 
\hline
$\Delta S_{nHB-nHB}^{i0-jk'} [10^{-2}k_B]$ & 2.2 & 0.6 & 4.3 & 1.1 &
5.6 & 1.3 & 4.1 & 0.9 \\

$\Delta S_{nHB-nHB}^{i0-jk'}/S_{nHB-nHB}^{i0-jk;B} [\%]$ & 149 & 41 &
291 & 74 & 378 & 88 & 277 & 61 \\
\hline
$\Delta S_{nHB-nHB}^{i0-jk'*} [10^{-2}k_B]$ & 0.9 & 0.3 & 1.3 & 0.3 &
1.6 & 0.4 & 1.2 & 0.3 \\

$\Delta S_{nHB-nHB}^{i0-jk'*}/S_{nHB-nHB}^{i0-jk*;B} [\%]$ & 1216 & 405 &
1757 & 405 & 2162 & 541 & 1622 & 405 \\
\hline
$\Delta S_{HB-nHB}^{i0-jk'} [10^{-2}k_B]$ & 0.7 & 0.2 & 0.9 & 0.2 & 1.2
& 0.3 & 0.9 & 0.2 \\

$\Delta S_{HB-nHB}^{i0-jk'}/S_{HB-nHB}^{i0-jk;B} [\%]$ & 875 & 250 &
1125 & 250 & 1500 & 375 & 1125 & 250 \\
\hline
$\Delta S_{HB-nHB}^{i0-jk'*} [10^{-2}k_B]$ & 0.8 & 0.2 & 0.9 & 0.2 & 1.2
& 0.3 & 0.9 & 0.3 \\

$\Delta S_{HB-nHB}^{i0-jk'*}/S_{HB-nHB}^{i0-jk*;B} [\%]$ & 145 & 26 &
164 & 26 & 218 & 55 & 164 & 55 \\
\hline
\end{tabular}
\end{center}
\label{t1}
\end{table*}